\documentclass[prl,twoside,twocolumn,showpacs,superscriptaddress,floatfix]{revtex4}
\usepackage{times,graphicx,bbm,amsmath,amssymb}
\usepackage{epsfig,color}
\newcommand{\bmsigma}{\boldsymbol \sigma} 
\newcommand{\bmLambda}{\boldsymbol \Lambda}

\def\S{{\sf S}}
\newcommand{\bmlambda}{\boldsymbol \lambda} \def\X{\boldsymbol{X}}
\def\Tr{\hbox{Tr}} \def\sigmaCM{\boldsymbol{\sigma}}
\begin{document}
\title{Quantifying the non-Gaussian character of a quantum state by
quantum relative entropy}
\author{Marco G. Genoni}
\affiliation{Dipartimento di Fisica dell'Universit\`a di Milano,
I-20133, Milano, Italia}
\affiliation{CNISM, UdR Milano Universit\`a, I-20133 Milano, Italia}
\author{Matteo G. A. Paris}
\affiliation{Dipartimento di Fisica dell'Universit\`a di Milano,
I-20133, Milano, Italia}
\affiliation{CNISM, UdR Milano Universit\`a, I-20133 Milano, Italia}
\affiliation{ISI Foundation, I-10133 Torino, Italia}
\author{Konrad Banaszek}
\affiliation{Institute of Physics, Nicolaus Copernicus University,
PL-87-100 Toru\'{n}, Poland}
\date{\today}
\begin{abstract}
We introduce a novel measure to quantify the non-Gaussian character of a
quantum state: the quantum relative entropy between the state under
examination and a reference Gaussian state. We analyze in details the
properties of our measure and illustrate its relationships with relevant
quantities in quantum information as the Holevo bound and the
conditional entropy; in particular a necessary condition for the
Gaussian character of a quantum channel is also derived.  The evolution
of non-Gaussianity (nonG) is analyzed for quantum states undergoing
conditional Gaussification towards twin-beam and de-Gaussification
driven by Kerr interaction. Our analysis allows to assess nonG as a
resource for quantum information and, in turn, to evaluate the
performances of Gaussification and de-Gaussification protocols.
\end{abstract}
\maketitle
\par{\em Introduction}---The use of Gaussian states and operations
allows the implementation of relevant quantum information protocols
including teleportation, dense coding and quantum cloning \cite{Brau}.
Indeed, the Gaussian sector of the Hilbert space plays a crucial role in
quantum information processing with continuous variables (CV),
especially for what concerns quantum optical implementations \cite{GRG}.
On the other hand, quantum information protocols required for long
distance communication, as for example entanglement distillation and
entanglement swapping, require nonG operations \cite{NoGO}.  Besides, it
has been demonstrated that using nonG states and operations
teleportation \cite{Tom,IPS2a,IPS2b} and cloning \cite{nonGclon} of
quantum states may be improved. Indeed, de-Gaussification protocols for
single-mode and two-mode states have been proposed
\cite{Tom,IPS2a,IPS2b,IPS1,KorolkovaKerr} and realized
\cite{IPS_Wenger}.  From a more theoretical point of view, it should be
noticed that any strongly superadditive and continuous functional is
minimized, at fixed covariance matrix (CM), by Gaussian states.  This is
crucial to prove extremality of Gaussian states and Gaussian operations
\cite{Wolf1,Wolf2} for various quantities such  as channel capacities
\cite{HW01}, multipartite entanglement measures \cite{EM} and
distillable secret key in quantum key distribution protocols.  Overall,
nonG appears to be a resource for CV quantum information and a question
naturally arises on whether a convenient measure to quantify the nonG
character of a quantum state may be introduced.  Notice that the notion
of nonG already appeared in classical statistics in the framework of
independent component analysis \cite{ICA}.
\par
The first measure of nonG of a CV state $\varrho$ has been
suggested in \cite{nonG_HS} based on the Hilbert-Schmidt (HS) distance
between $\varrho$ and a reference  Gaussian state.  In turn, the
HS-based measure has been used to characterize the role of
nonG as a resource for teleportation \cite{Illum1,Illum2}
and in promiscuous quantum correlations in CV systems \cite{Illum3}.
Here we introduce a novel measure
$\delta[\varrho]$ based on the quantum relative entropy between
$\varrho$ and a reference Gaussian state. The novel quantity is related
to information measures and allows to assess nonG as a
resource for quantum information as well as the performances of
Gaussification and de-Gaussification protocols.  In the following, after
introducing its formal definition and showing that it can be easily
computed for any state, either single-mode or multi-mode, we analyze in
details the properties of $\delta[\varrho]$  as well as its dynamics
under Gaussification \cite{PlenioGauss} and de-Gaussification
protocols.
\par
{\em Gaussian states}---Let us consider a CV system made of $d$ bosonic modes
described by the mode operators $a_k$, $k=1\dots d$, with
commutation relations $[a_k,a_j^{\dag}]=\delta_{kj}$. A quantum
state $\varrho$ of $d$ bosonic modes is fully described by its
characteristic function $\chi[\varrho](\bmlambda) = \Tr[\varrho\:D(\bmlambda)]$
where $D(\bmlambda) = \bigotimes_{k=1}^d D_k(\lambda_k)$
is the $d$-mode displacement operator, with $\bmlambda =
(\lambda_1,\dots,\lambda_d)^T$, $\lambda_k \in \mathbbm{C}$, and where
$D_k(\lambda_k) =\exp\{\lambda_k a_k^{\dag} - \lambda_k^* a_k \}$ is the
single-mode displacement operator.
The canonical operators are given by
$q_k = (a_k + a^{\dag}_k)/\sqrt{2}$ and
$p_k = (a_k - a_k^{\dag})/\sqrt{2}i$
with commutation relations given by $[q_j,p_k]=i\delta_{jk}$.
Upon introducing the vector $\boldsymbol{R}=(q_1,p_1,\dots,q_d,p_d)^T$,
the CM $\bmsigma\equiv\bmsigma[\varrho]$ and the vector
of mean values $\X \equiv \X[\varrho]$ of a quantum state
$\varrho$ are defined as
$\sigma_{kj} =\frac{1}{2}\langle R_k R_j + R_j R_k\rangle - \langle R_j
\rangle\langle R_k\rangle$ and
${X}_j = \langle R_j \rangle$,
where
$\langle O \rangle = \Tr[\varrho\:O]$ is the expectation value
of the operator $O$.
A quantum state
$\varrho_G$
is said to be Gaussian if
its characteristic function is Gaussian, that is
$\chi[\varrho_G](\bmLambda) = \exp \left\{- \frac{1}{2}
\bmLambda^T \bmsigma \bmLambda + \X^T\bmLambda \right\}
$,
where $\bmLambda$ is the real vector $\bmLambda = (\hbox{Re}
\lambda_1, \hbox{Im}\lambda_1, \dots, \hbox{Re} \lambda_d,
\hbox{Im}\lambda_d)^T$. Once the CM and
the vectors of mean values are given, a Gaussian state is fully
determined.  For a system
of $d$ bosonic modes the most general Gaussian state is described by $d(2d +3)$
independent parameters.
\par
{\em Non-Gaussianity of a CV state}---The von Neumann entropy of a
quantum state is defined as $\S(\varrho) = - \Tr[ \varrho \: \log \varrho ]$.
The von Neumann entropy is non-negative and equals zero iff $\varrho$ is
a pure state. In order to quantify the nonG character of a quantum
state $\varrho$ we employ the quantum relative entropy (QRE)
$\S(\varrho \rVert \tau)=\Tr[\varrho (\log \varrho - \log\tau)]$
between $\varrho$ and a reference Gaussian state $\tau$.
As for its classical counterpart, the Kullback-Leibler divergence, it can
be demonstrated that $0\leq \S(\varrho \rVert \tau) < \infty$ when it
is definite, \emph{i.e.} when
$\mathrm{supp} \: \varrho \subseteq \: \mathrm{supp} \: \tau$.
In particular $\S(\varrho \rVert \tau)=0$ iff $\varrho\equiv\tau$.
This quantity, though not defining  a proper metric in the Hilbert
space, has been widely used in different fields of quantum information
as a measure of statistical distinguishability for quantum states
\cite{SchumacherRelEnt,VedralRelEnt}.
Therefore, given a quantum state $\varrho$ with finite
first and second moments, we define its nonG as
$\delta[\varrho] =  \S(\varrho\rVert \tau)$,
where the reference state $\tau$ is
is the Gaussian state with
 $\X[\varrho] = \X[\tau]$ and
$\boldsymbol{\sigma}[\varrho] = \boldsymbol{\sigma}[\tau] $,
\emph{i.e.} the Gaussian state with the same CM
$\bmsigma$ and the same vector $\X$ of the state $\varrho$.
Finally, since $\tau$ is Gaussian, then $\log\tau$ is a
polynomial operator of the second order in the canonical variables which,
together with the fact that $\tau$ and $\rho$ have the same CM
leads to $\Tr[(\tau -\varrho)\: \log\tau]=0$ \cite{HolevoVNEnt}, {\em
i.e.} $\S(\varrho\rVert\tau) = \S(\tau) - \S(\varrho)$. Thus we have
\begin{align}
\delta[\varrho] = \S(\tau) - \S(\varrho)
\label{eq:nonG2}
\end{align}
\emph{i.e.} nonG is the difference between the von Neumann
entropies of $\tau$ and $\varrho$. In turn, several properties
of the nonG measure $\delta [\varrho ]$ may be derived from the
fundamental properties of QRE \cite{VedralRelEnt,
SchumacherRelEnt}.  In the following we summarize the relevant ones
by the following Lemmas:
\par\noindent
{\bf L1}: $\delta[\varrho]$ is a well defined non negative quantity,
that is $0\leq\delta[\varrho] < \infty$ and $\delta[\varrho]=0$ iff
$\varrho$ is a Gaussian state.  {\bf Proof}: Nonnegativity is guaranteed
by the nonnegativity of the quantum relative entropy. Moreover, if
$\delta[\varrho]=0$ then $\varrho=\tau$ and thus it is a Gaussian state.
If $\varrho$ is a Gaussian state, then it is uniquely identified by its
first and second moments and thus the reference Gaussian state $\tau$ is
given by $\tau=\varrho$, which, in turn, leads to
$\delta[\varrho]=\S(\varrho \rVert \tau)=0$.
\par\noindent
{\bf L2}:  $\delta[\varrho]$ is a continuous functional.
{\bf Proof}: It follows from the continuity
of trace operation and QRE.
\par\noindent
{\bf L3}: $\delta[\varrho]$ is additive for factorized states:
$\delta[\varrho_1\otimes\varrho_2] = \delta[\varrho_1] +
\delta[\varrho_2]$.
As a corollary we have that if $\varrho_2$
is a Gaussian state, then $\delta[\varrho]=\delta[\varrho_1]$.
{\bf Proof}: The overall reference
Gaussian state is the tensor product of the relative reference
Gaussian states, $\tau=\tau_1\otimes\tau_2$. The lemma thus follows from
the additivity of QRE and the corollary from {L1}.
\par\noindent
{\bf L4}:  For a set of states $\{\varrho_k\}$
having the same first and second moments, then
nonG is a convex functional,
that is $\delta[\sum_k p_k \varrho_k]\leq \sum_k p_k \delta[\varrho_i]$,
with $\sum_k p_k =1$.
{\bf Proof}: The states $\varrho_k$, having the same
first and second moments, have the same reference Gaussian
state $\tau$ which in turn is the reference Gaussian state
of the convex combination $\varrho=\sum_k p_k \varrho_k$.
Since conditional entropy $\S(\varrho\rVert\tau)$
is a jointly convex functional respect to both states, we have
$\delta[\sum_k p_k \varrho_k]=\S(\sum_k p_k \varrho_k \rVert \tau)
\leq \sum_k p_k \S(\varrho_k \rVert \tau) = \sum_k p_k \delta[\varrho_k]$.
$\square$ Notice that, in general, nonG is not convex, as it may easily
proved upon considering the convex combination of two Gaussian states
with different parameters.
\par\noindent
{\bf L5}:  If $U_{b}$ is a unitary evolution corresponding to a symplectic
transformation in the phase space, \emph{i.e.} if $U_{b}=\exp\{-i H\}$ with
$H$ at most bilinear in the field operator, then
$\delta[U_{b}\varrho U_{b}^{\dag}] = \delta[\varrho]$.
{\bf Proof}:  Let us consider $\varrho^\prime = U_{b}\varrho U_{b}^\dag$,
where $U$ is at most bilinear in the field-mode, then its
CM transforms as $\sigmaCM[\varrho^\prime] =
\Sigma \sigmaCM[\varrho] \Sigma^T$, $\Sigma$ being the symplectic
transformation associated to $U$. At the same time the vector of
mean values simply translates to $\X^{\prime}=\X+\X_0$. Since any
Gaussian state is fully characterized by its first and second
moments, then the reference state must necessarily transform as
$\tau^\prime = U_{b} \tau U_{b}^\dag$, \emph{i.e.} with the same unitary
transformation $U$. Lemma follows from invariance of QRE under
unitary transformations.
\par\noindent
{\bf L6}: nonG is monotonically decreasing
 under partial trace, that is
$\delta[\Tr_B [\varrho]] \leq \delta[\varrho]$.
{\bf Proof}: Let us consider $\varrho^\prime = \Tr_B [\varrho]$. Its
CM is the submatrix of $\sigmaCM[\varrho]$ and its first
moment vector is the \emph{subvector} of $\X[\varrho]$ corresponding
to the relevant Hilbert space. As before, also the new reference Gaussian
state must necessarily transform as $\tau^\prime = \Tr_B [\tau]$. QRE
is monotonous decreasing under partial trace and the
lemma is proved.
\par\noindent
{\bf L7}: nonG is monotonically decreasing
 under Gaussian quantum channels, that is
$\delta[\mathcal{E}_G(\varrho)] \leq \delta[\varrho]$.
{\bf Proof}: Any Gaussian quantum channel can be written as
$\mathcal{E}_G(\varrho) =
\Tr_E [ U_{b} (\varrho\otimes\tau_E) U^{\dag}_{b}]$,
where $U_{b}$ is a unitary operation corresponding to an Hamiltonian
at most bilinear in the field modes and where $\tau_E$ is a Gaussian state
\cite{GaussChannel}. Then, by using lemmas {L3}, {L5} and
{L6} we obtain
$\delta[\mathcal{E}_G(\varrho)] \leq
\delta[U_{b} (\varrho\otimes\tau_E) U^{\dag}_{b}] = \delta[\varrho]$.
$\square$ In turn, {L7} provides a necessary condition for a
channel to be Gaussian: given a quantum channel $\mathcal{E}$, and a
generic quantum state $\varrho$, if the inequality
$\delta[\mathcal{E}(\varrho)]\leq \delta[\varrho]$ is not fulfilled, the
channel is nonG.
\par
{\em Maximally non-Gaussian states}---Let us now consider a single mode
($d=1$) system and look for states with the maximum amount of
nonG at fixed average number of photons $N=\langle a^{\dag}a
\rangle$.
Since $\delta[\varrho] = \S(\tau) - \S(\varrho)$ we have to
maximize $\S(\tau)$ and, at the same time, minimize $\S(\varrho)$.
For a single-mode system the most general
Gaussian state can be written as $\varrho_G= D(\alpha) S(\zeta)
\nu(n_t) S^\dag (\zeta) D^\dag(\alpha)$, $D(\alpha)$ being the
displacement operator, $S(\zeta) = \exp(\frac12 \zeta a^{\dag 2}
- \frac12 \zeta^* a^2)$ the squeezing operator, $\alpha,\zeta \in
{\mathbbm C}$, and $\nu(n_t)=(1+n_t)^{-1} [n_t/(1+n_t)]^{a^\dag
a}$ a thermal state with $n_t$ average number of photons.
Displacement and squeezing applied to thermal states increase the
overall energy, while entropy is an increasing monotonous function of
the number of thermal photons $n_t$ and is invariant under
unitary operations, thus, at fixed energy, $\S(\tau)$ is maximized for
$\tau=\nu(N)$. Therefore, the state with the maximum amount of
nonG must be a pure state (in order to have $\S(\varrho)=0$)
with the same CM $\bmsigma=(N+\frac12) {\mathbbm I}$
of the thermal state $\nu(N)$.  These
properties individuate the superpositions of Fock states
$|\psi_{N}\rangle = \sum_k \alpha_k |n + l_k\rangle$ where $n\geq 0$,
$l_k\geq l_{k-1} + 3$ or $l_k=0$, with the constraint 
$N=\langle a^{\dag}a\rangle$,
{\em i.e} $n+\sum_k |\alpha_k |^2 l_k = N = (\det
\bmsigma[\nu(N)])^{\frac12} -\frac12$.
These represent maximally
nonG states, and include Fock states
$|\psi_{N}\rangle=|N\rangle$ as a special
case.  Let us consider now $d$-mode quantum states with fixed average
number of photons $\sum_{k=1}^d\Tr[a_k^{\dag}a_k\varrho]=N=\sum_{k} n_k$.
Also in this case maximally nonG states are pure
states; the CM being equal to that of a multimode classical
state $\tau=R\, \otimes_{k}\nu(m_k) R^{\dag}$, $\sum_{k}m_k=N$,
where we denote with $R$ a generic set of symplectic passive operations
(\emph{e.g.} beam splitter evolution) which do not increase the energy.
In order to maximize $\S(\tau)=\sum_{k}
\S(\nu(m_k))$ we have to
choose $m_k=N/d$ for every $k$.  As for example,
factorized states of the form
$|\Psi_{N}\rangle =  |\psi_{N/d}\rangle^{\otimes d} $,
whose reference Gaussian states are
$\tau=[\nu(N/d)]^{\otimes d}$,
are maximally nonG states at fixed $N$.
Of course for the multi-mode case there are other more complicated
classes of maximally nonG states that include also entangled pure
states. Finally, we observe that the maximum value of nonG is
a monotonous increasing function of the number of photons $N$.
\par
{\em Non-Gaussianity in quantum information}---Gaussian states are
extremal for several functionals
in quantum information \cite{Wolf1}. In
the following we consider two relevant examples, and show how
extremality properties may be quantified by the nonG measure
$\delta[\varrho]$.  Let us first consider a generic communication
channel where the letters from an alphabet are encoded onto a set of
quantum states $\varrho_k$ with probabilities $p_k$.  The \emph{Holevo
Bound} represents the upper bound to the accessible information, and is
defined as $\chi(\varrho) = \S(\varrho) - \sum_k p_k\: \S(\varrho_k)$
where $\varrho= \sum_k p_k \varrho_k$ is the overall ensemble sent
through the channel.  Upon fixing the CM (and the first
moments) of $\varrho$ we rewrite the Holevo bound as $\chi(\varrho) =
\S(\tau) - \delta[\varrho] - \sum_k p_k\: \S(\varrho_k)$, where $\tau$ is
the Gaussian reference of $\varrho$.  This highlights the role
of the nonG $\delta[\varrho]$ of the overall state in
determining the amount of accessible information: at fixed CM
the most convenient encoding corresponds to a set of pure states
$\varrho_k$, $\S(\varrho_k)=0$, forming an overall Gaussian ensemble
with the largest entropy. In other words, at fixed CM, we
achieve the maximum value of $\chi$ upon encoding encoding symbols onto
the eigenstates of the corresponding Gaussian state \cite{dru}.
If the alphabet is encoded onto the eigenstates
of a given state $\varrho$, we have
$\chi(\varrho)= \S(\tau) - \delta[\varrho]$. This suggests an
operational interpretation of nonG $\delta[\varrho]$ as the
loss of information we get by encoding symbols on the eigenstates of
$\varrho$ rather than on those of its reference Gaussian
state.
\par
Let us now consider the state $\varrho_{AB}$ describing two quantum
systems $A$ and $B$ and define the conditional entropy $\S(A \vert B ) =
\S(\varrho_{AB}) - \S(\varrho_B)$.  Let us fix the CM of
$\varrho_{AB}$ and thus also that of $\varrho_B$, and consider the
reference Gaussian states $\tau_{AB}$ and $\tau_{B}$. We may write $\S(A
\vert B) = \S_G(A \vert B) - ( \delta[\varrho_{AB}] - \delta[\varrho_B]
)$ where $\S_G(A \vert B)=\S(\tau_{AB})-\S(\tau_B)$, \emph{i.e.} the
conditional entropy evaluated for the reference Gaussian states
$\tau_{AB}$ and $\tau_B$.  Then, upon using {L6} we have
$\delta[\varrho_{AB}] - \delta[\varrho_B] \geq 0$ and thus
$\S(A\vert B) \leq \S_G(A \vert B)$, {\em i.e} the maximum of conditional
entropy at fixed CM is achieved by Gaussian states.  In
classical information theory the conditional entropy $H(X\vert Y) =
H(X,Y) - H(Y)$, where von Neumann entropies are replaced by Shannon
entropies of classical probability distributions, is a positive quantity
and may be interpreted \cite{SlepWolf} as the amount of partial
information that Alice must send to Bob so that he gains full knowledge
of $X$ given his previous knowledge from $Y$.
When quantum systems are involved the conditional entropy may be
negative, negativity being a sufficient condition for the entanglement
of the overall state $\varrho_{AB}$. This \emph{negative} information
may be seen as follows \cite{HoroPartial} for a discrete variable
quantum system. Given an unknown quantum state
distributed over two systems, we can discriminate between
two different cases: if
$\S(A\vert B)\geq 0$, as in the classical case, it gives the amount
of information that Alice should send to Bob to give him the full
knowledge of the overall state $\varrho_{AB}$.  When $\S(A\vert B) < 0$
Alice does not need to send any information to Bob and moreover they
gain $- \S(A\vert B)$ ebits.  If we conjecture that this interpretation
can be extended to the CV case, the relation $\S(A\vert B) \leq
\S_G(A \vert B)$ ensures that, at fixed CM, nonG states
always perform better: Alice needs to send less
information, or, for negative values of the conditional entropy, more
entanglement is gained.  Moreover, since negativity of conditional
entropy is a sufficient condition for entanglement \cite{CerfAdami} we
have that for any given bipartite quantum state $\varrho_{AB}$, if the
conditional entropy of the reference Gaussian state $\tau_{AB}$ is
negative, then $\varrho_{AB}$ is an entangled state. Though being a
weaker condition than the negativity of
$\S(A|B)$, this is a simple and easy computable test for entanglement which
is equivalent to evaluate
the symplectic eigenvalues \cite{SerafiniGauss} of the involved
Gaussian states.
\par
{\em Gaussification and de-Gaussification protocols}---Since the
amount of nonG of a state affects its performances in
quantum information protocols a question naturally arises on whether
this may be engineered or modified at will. As concerns Gaussification,
Lemma {L7} assures that Gaussian maps do not increase
nonG. In turn, the simplest example of Gaussification map
is provided by dissipation in a thermal bath \cite{nonG_HS}, which
follows from bilinear interactions between the systems under
investigation and the environment. On the other hand, a conditional
iterative Gaussification protocol has been recently proposed
\cite{PlenioGauss} which cannnot be reduced to a
trace-preserving Gaussian quantum map.
It requires only the use of passive elements and on/off photodetectors.
Given a bipartite pure state in the Schmidt form
$|\psi^{(k)}\rangle = \sum_{n=0}^{\infty} \alpha^{(k)}_{n,n}|n,n\rangle$
the state at $k+1$-th step of the protocol has the same Schmidt form
with $\alpha^{(k+1)}_{n,n} = 2^{-n} \sum_{r=0}^n \binom{n}{r}
\alpha^{(k)}_{r,r} \alpha^{(i)}_{n-r,n-r} $. We have considered
the initial nonG superposition $|\psi^{(0)}\rangle =
(1+\lambda^2)^{-1/2}(|0,0\rangle + \lambda |1,1\rangle)$ which is
asymptotically driven towards the Gaussian twin-beam state
$|\psi\rangle = \sqrt{1-\lambda^2}
\sum_{n=0}^{\infty} \lambda^n |n,n\rangle$. We have evaluated
nonG at any step of the protocol,
for every value of $\lambda$.
\begin{figure}[h!]
\includegraphics[width=0.21\textwidth]{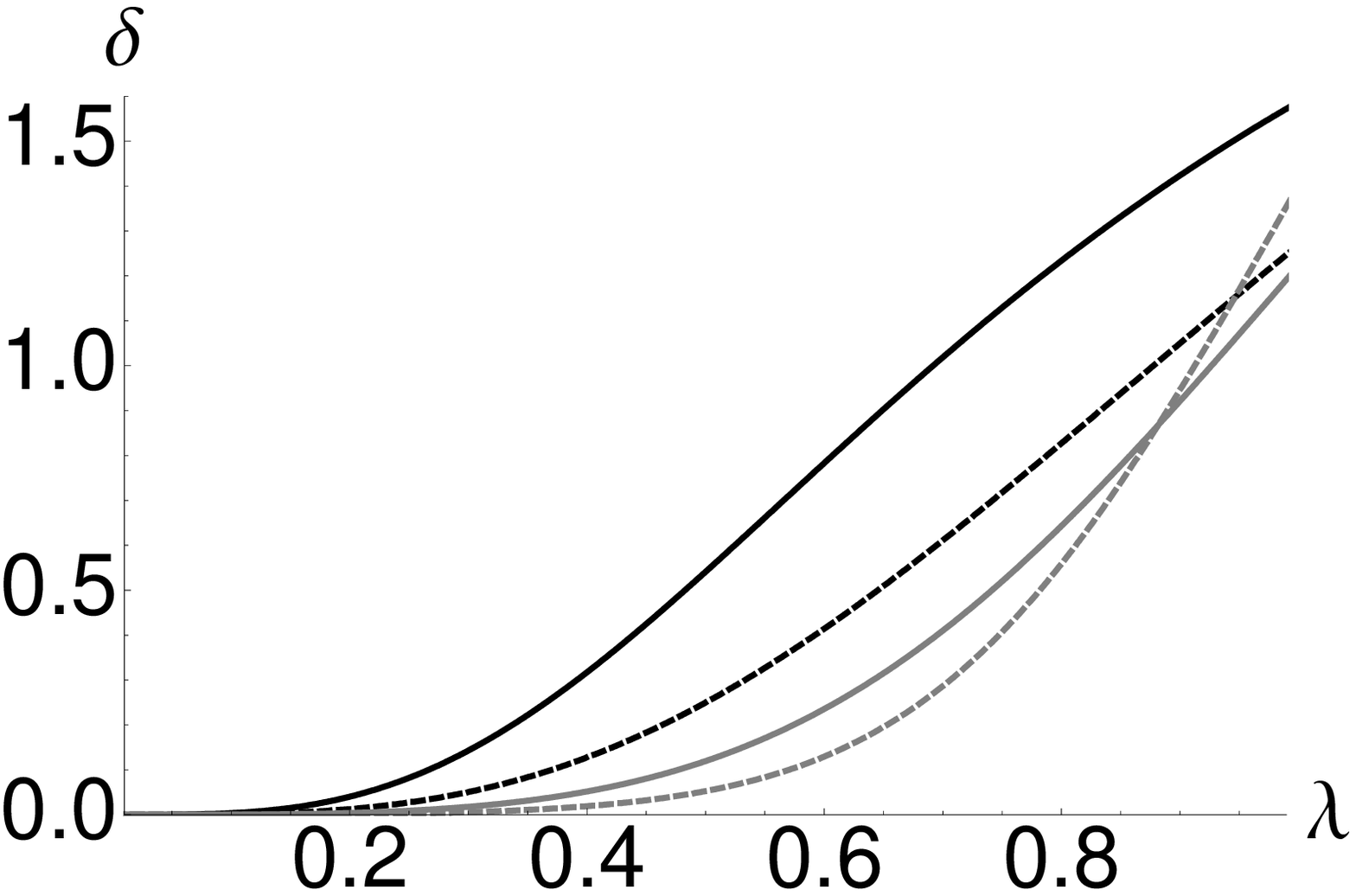}$\:$
\includegraphics[width=0.21\textwidth]{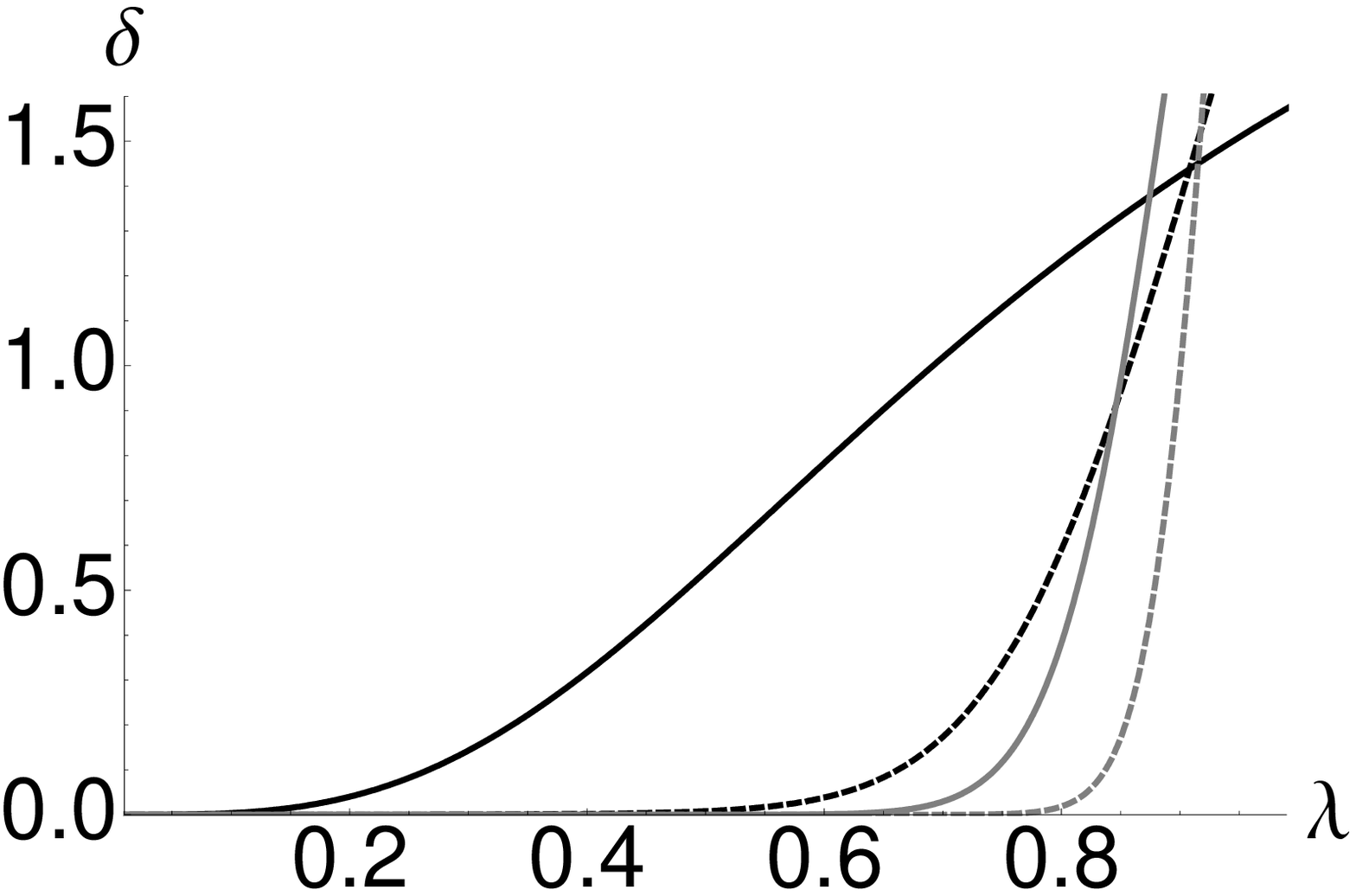}
\caption{
\label{f:fig}
(Left): nonG after some steps of the conditional Gaussification
protocol of Ref. \cite{PlenioGauss} considering as the initial state
the nonG superposition $|\psi^{(0)}\rangle =
(1+\lambda^2)^{-1/2}(|0,0\rangle + \lambda |1,1\rangle)$:
black-solid: initial state; black-dashed
: step 1; gray-solid: step 2; gray-dashed: step 3.
(Right): black-solid: initial state; black-dashed
: step 5; gray-solid : step 10; gray-dashed: step 20.}
\end{figure} \\
Results
are reported in Fig. \ref{f:fig}: for the first steps,
nonG decreases monotonically for almost all
values of $\lambda$ (only at the third step, for $\lambda\approx 1$
the state is more nonG than at the previous steps). Notice that
increasing the number of steps
nonG may also increase, {\em e.g}, for $\lambda\approx 1$,
$\delta$ reaches very high values and the maximum value increases.
On the other hand, the overall effectiveness of the protocol is
confirmed by our analysis, since the range of values of $\lambda$ for
which $\delta\approx 0$ increases at each step of the protocol. In other
words, though not being a proper Gaussian map, the conditional protocol
of \cite{PlenioGauss} indeed provides an effective Gaussification
procedure.
\par
\begin{figure}[h!]
\includegraphics[width=0.33\textwidth]{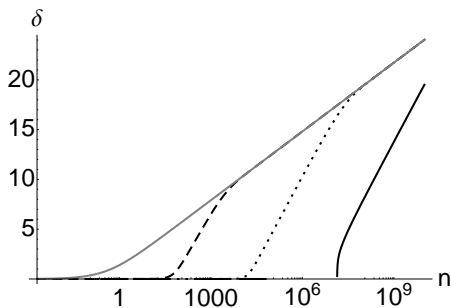}
\caption{
\label{f:f2}
NonG of coherent states undergoing Kerr interaction as
a function of the average number of photons and for different values of the
coupling  constant $\gamma$. Black dashed line: $\gamma = 10^{-2}$; black
dotted: $\gamma = 10^{-4}$; solid black: $\gamma = 10^{-6}$.
The gray solid lines is the maximum nonG at fixed number of photons.}
\end{figure}

Conditional de-Gaussification procedures have been recently proposed and
demonstrated \cite{IPS2a,IPS2b,IPS1,IPS_Wenger}. Here we rather consider
the unitary de-Gaussification evolution provided by self-Kerr
interaction $U_{\gamma}= \exp \{ -i \gamma (a^{\dag}a)^2 \}$
\cite{LeuchsKerr,AndersenKerr}, which does not correspond
to a symplectic transformation and leads to a nonG state even if
applied to a Gaussian state.
We have evaluated the nonG of the state obtained from a
coherent state $|\alpha\rangle$
undergoing Kerr interaction. Results are reported in Fig. \ref{f:f2}
as a function of the average number of photons (up to $10^9$ photons)
and for different values of the coupling constant $\gamma$ .
As it is apparent from the plot,
nonG is an increasing function of the number
of photons and the Kerr coupling $\gamma$. For $\gamma\approx 10^{-2}$,
the maximum nonG achievable at fixed energy is
quite rapidly achieved. For more realistic values of the nonlinear
coupling, {\em i.e.} $\gamma\leq 10^{-6}$ nonG states
may be obtained only for a large average number of photons
in the output state. In fact, to obtain entanglement,
experimental realizations \cite{LeuchsKerr,AndersenKerr}
involve pulses with an average
number of the order of $10^8$ photons, which are needed to compensate
the almost vanishing small Kerr nonlinearities of standard glass fiber.
We finally notice that a good measure for the nonG character of quantum
states allows us to define a measure of the nonG character
of a quantum operations. Let us denote by ${\cal G}$ the whole
set of Gaussian states. A convenient definition for the
nonG of of a map ${\cal E}$ reads as
$\delta[{\cal E}] = \max_{\varrho \in {\cal G}} \delta [{\cal E}
(\varrho)]$, where ${\cal E} (\varrho)$ denotes the quantum state
obtained after the evolution imposed by the map.
\par{\em Conclusions}---We have introduced a novel measure
to quantify the nonG character of a CV quantum state based on
quantum relative entropy.  We have analyzed in details the properties
owned by this measure and its relation with some relevant quantities in
quantum information. In particular, a necessary condition for the
Gaussian character of a quantum channel and a sufficient condition for
entanglement of bipartite quantum states can be derived. Our measure
is easily computable for any CV state and allows to assess nonG as a
resource for quantum technology. In turn, we exploited our measure to
evaluate the performances of conditional Gaussification towards
twin-beam and de-Gaussification processes driven by Kerr interaction.
\par
We thank M. Horodecki for discussions on conditional entropy in the
CV regime. This work has been supported by the CNISM-CNR convention,
Polish budget funds for scientific research projects 
(Grant No. N N202 1489 33), and the European Commission under the 
Integrated Project QAP (Contract No. 015848).


\begin{thebibliography}{99}
\bibitem{Brau} S. L. Braunstein et al.,
Rev. Mod. Phys {\bf 77}, 513 (2005).
\bibitem{GRG}
J. Eisert et al., Int. J. Quant. Inf. {\bf 1}, 479 (2003);
A. Ferraro et al., {\em Gaussian States in Quantum Information},
(Bibliopolis, Napoli, 2005); F. Dell'Anno et al., Phys. Rep. {\bf
428}, 53 (2006).
\bibitem{NoGO} J. Eisert et al., Phys. Rev. Lett. {\bf 89}, 137903 (2002);
G. Giedke et al., Phys. Rev. A {\bf 66}, 032316 (2002).
\bibitem{Tom} T. Opatrny et al., Phys.Rev. A {\bf
61}, 032302 (2000).
\bibitem{IPS2a} P. T. Cochrane et al., Phys Rev. A {\bf 65}, 062306 (2002).
\bibitem{IPS2b} S. Olivares et al., Phys. Rev. A {\bf 67}, 032314
(2003).
\bibitem{nonGclon} N. J. Cerf et al., Phys. Rev. Lett. {\bf 95}, 070501 (2005).
\bibitem{IPS1} S. Olivares et al., J. Opt. B {\bf 7}, S392 (2005).
\bibitem{KorolkovaKerr} T. Tyc at al., New. J. Phys. {\bf 10}, 023041 (2008).
\bibitem{IPS_Wenger} J. Wenger et al,
Phys. Rev. Lett. {\bf 92}, 153601 (2004).
\bibitem{Wolf1} M. M. Wolf et al., Phys. Rev. Lett. {\bf 96}, 080502 (2006).
\bibitem{Wolf2} M. M. Wolf et al, Phys. Rev. Lett. {\bf 98}, 130501 (2007).
\bibitem{HW01} A. S. Holevo et al., Phys. Rev. A {\bf 63}, 032312
(2001).
\bibitem{EM} L. M. Duan at al, Phys. Rev. Lett. {\bf 84}, 4002 (2000);
\bibitem{ICA} A. Hyvarinen et al., {\em Independent Component Analysis},
(John Wiley \& Sons, 2001).
\bibitem{nonG_HS} M. G. Genoni et al. Phys. Rev. A {\bf 76}, 042327 (2007).
\bibitem{Illum1} F. Dell'Anno et al., Phys. Rev. A {\bf 76}, 022301 (2007).
\bibitem{Illum2} F. Dell'Anno et al., arXiv:0710.3259v2.
\bibitem{Illum3} G. Adesso et al., Phys. Rev. A {\bf 76}, 022315 (2007).
\bibitem{VedralRelEnt} V. Vedral, Rev. Mod. Phys. {\bf 74}, 197--234 (2002).
\bibitem{SchumacherRelEnt} B. Schumacher et al., ArXiv: quant-ph/0004045.
\bibitem{HolevoVNEnt} A. S. Holevo et al., Phys. Rev. A {\bf 59}, 1820 (1999).
\bibitem{GaussChannel} J. Eisert et al. , ArXiv:quant-ph/0505151.
\bibitem{dru} C. M. Caves et al., Rev. Mod. Phys. \textbf{66}, 481 (1994)
\bibitem{SlepWolf} D. Slepian and J. Wolf, IEEE Trans. Inf. Theory {\bf 19}, 147
(1971).
\bibitem{HoroPartial} M. Horodecki et al., Nature {\bf 436}, 673 (2005).
\bibitem{CerfAdami} N. J. Cerf et al., Phys. Rev. A {\bf 60}, 893 (1999).
\bibitem{SerafiniGauss} A. Serafini et al., J. Phys. B {\bf 37}, L21 (2004).
\bibitem{PlenioGauss} D. Browne et al., Phys. Rev. A {\bf 67}, 062320 (2003).
\bibitem{LeuchsKerr} C. Silberhorn et al., Phys. Rev. Lett. {\bf 86}, 4267 (2001).
\bibitem{AndersenKerr} O. Glockl et al., Phys. Rev. A {\bf 73}, 012306 (2006).
\end{thebibliography}
\end{document}